 \documentclass[apj]{emulateapj}
\usepackage{color}
\usepackage{epsfig}
\newcommand{\myemail}{guofan.ustc@gmail.com}

\doublespace

\shorttitle{Particle Transport in Magnetic Turbulence} \shortauthors{Guo and Giacalone}

\begin{document}

\title{Small-scale Gradients of Charged Particles in the Heliospheric Magnetic Field}

\author{Fan Guo \altaffilmark{1,2} and Joe Giacalone \altaffilmark{2}}

\altaffiltext{1}{Theoretical Division, Los Alamos National Laboratory, Los
Alamos, NM 87545}

\altaffiltext{2}{Department of Planetary Sciences and Lunar and Planetary
Laboratory, University of Arizona, 1629 E. University Blvd., Tucson, AZ 85721}

\email{\myemail}

\begin{abstract}
Using numerical simulations of charged-particles propagating in the heliospheric magnetic field, we study small-scale gradients, or ``dropouts", in the intensity of solar energetic particles seen at 1 AU. We use two turbulence models, the foot-point random motion model \citep{Jokipii1969,Giacalone2006} and two-component model \citep{Matthaeus1990}, to generate fluctuating magnetic fields similar to spacecraft observations at $1$ AU. The turbulence models include a Kolmogorov-like magnetic field power spectrum containing a broad range of spatial scales from those that lead to large-scale field-line random walk to small scales leading to resonant pitch-angle scattering of energetic particles. We release energetic protons ($20$ keV - $10$ MeV) from a spatially compact and instantaneous source. The trajectories of energetic charged particles in turbulent magnetic fields are numerically integrated. Spacecraft observations are mimicked by collecting particles in small windows when they pass the windows at a distance of $1$ AU. We show that small-scale gradients in the intensity of energetic particles and velocity dispersions observed by spacecraft can be reproduced using the foot-point random motion model. However, no dropouts are seen in simulations using the two-component magnetic turbulence model. We also show that particle scattering in the solar wind magnetic field needs to be infrequent for intensity dropouts to form.
\end{abstract}


\section{Introduction}
The propagation of energetic particles in turbulent magnetic fields is of great importance in space physics and astrophysics. The large-scale transport of charged particles is usually studied by solving the cosmic-ray transport equation first derived by E. N. Parker (\citeyear{Parker1965}). Determining diffusion coefficients in the equation is crucial to studying the transport of charged particles. The diffusion coefficients parallel and perpendicular to the background magnetic field ($\kappa_\parallel$ and $\kappa_\perp$) are considered to be quite different, with $\kappa_\parallel$ usually larger than $\kappa_\perp$ \citep{Jokipii1966,Giacalone1999}.

An unsolved issue in the transport of energetic particles is that the mean-free paths of energetic particles inferred from solar energetic particle (SEP) events are usually much longer than those derived from the quasi-linear theory \citep{Palmer1982,Bieber1994}. It was proposed that the anisotropy of magnetic turbulence may have influences on the diffusion coefficient parallel to the magnetic field \citep{Bieber1996,Chandran2000,Yan2002}, while some numerical simulations found that the diffusion coefficient only has a weak dependence on the anisotropy \citep{Giacalone1999,Qin2006}. The discrepancy between observations and theories is still not well resolved. 

The transport of charged particles normal to the magnetic field is also not well understood. Some analyses give a rather small cross-field diffusion with the ratio of the perpendicular diffusion coefficient to the parallel diffusion coefficient $\kappa_\perp/\kappa_\parallel \sim 10^{-4}$ or smaller \citep{Roelof1983}. Recent numerical simulations and analytical studies find a larger value of $\kappa_\perp/\kappa_\parallel \sim 0.02 $ - $0.05$ for energetic particles moving in the heliospheric magnetic field at $1$ AU \citep{Giacalone1999,Qin2002,Mattaeus2003}. However, some spacecraft measurements suggest that the ratio can reach $0.2$ or even larger \citep{Dwyer1997,Zhang2003}, which is unexpectedly large compared to those obtained from numerical simulations \citep{Giacalone1999}. Newly available data shows that impulsive SEP events are occasionally seen by all three spacecraft (STEREO A/B and ACE) with a separation of more than $100$ degrees in heliospheric longitude \citep{Wiedenbeck2010,Dresing2012,Wiedenbeck2013}. \citet{Giacalone2012} and \citet{Dresing2012} have suggested that $\kappa_\perp/\kappa_\parallel$ has to be as large as a few percent to explain these multi-spacecraft observations.

Impulsive SEPs are ideal for studying the transport of charged particles because their acceleration time is much shorter than the propagation time in the heliospheric magnetic field. Recently, observations by spacecraft such as ACE and \textit{Wind} have revealed new features of particle transport \citep{Mazur2000,Chollet2007,Chollet2008a,Chollet2008b,Chollet2011}. \citet{Mazur2000} reported that the intensity of impulsive SEP events often shows small-scale sharp gradients, or ``dropouts". These dropouts are commonly seen in impulsive SEP events and the typical convected distance between dropouts is about $0.03$ AU, similar to the correlation scale in the solar wind turbulence \citep[e.g.,][]{Wicks2010}. The occurrence of dropouts does not seem to be associated with rapid magnetic field changes, meaning that it is more related to some large-scale transport effects. The phenomenon indicates that the diffusion of energetic particles transverse to the local magnetic field is very small \citep{Chollet2011}; the transport of energetic particles in the solar wind is likely dominated by some other effect.
It is worth noting that the motions of energetic charged particles transverse to the magnetic field can be considered as two parts: $1$. particle motions across the local magnetic field due to drift or scattering; and $2$. particle motions along meandering magnetic field lines but normal to the mean magnetic field. The observed SEP dropouts may be interpreted as that the motions of particles across the local magnetic field is small, but a large part of perpendicular diffusion can be contributed by field-line random walk. These new observations have provided an excellent opportunity to examine and constrain relative contributions from these two effects to large-scale perpendicular transport.

Using numerical simulations that consider large-scale turbulent magnetic fields, \citet{Giacalone2000} have demonstrated that dropouts can be reproduced when energetic particles are released in a small source region near the Sun. This model is consistent with magnetic turbulence models that allow a large perpendicular diffusion coefficient due to field-line random walk \citep{Giacalone1999}. The time duration between numerically produced dropouts is several hours, which is similar to that observed in impulsive SEP events \citep{Mazur2000,Chollet2008b}. It also naturally reproduces the feature that the typical convected distance between dropouts is similar to the correlation scale in the solar wind turbulence \citep{Mazur2000}. \citet{Ruffolo2003} and \citet{Chuychai2007} have proposed a different idea based on the two-component turbulence model \citep{Matthaeus1990,Bieber1996}. They argue that some magnetic field lines in the solar wind can have a restricted transverse displacement. The corresponding magnetic flux tubes connecting to SEP source regions are concentrated by energetic particles. For magnetic field lines that meander in space, energetic particles that are initially confined to the field lines will diffuse away. However, this effect may depend on the two-component magnetic field model they use (a combination of a two-dimensional fluctuation and a one-dimensional fluctuation). \citet{Chuychai2007} studied the trapping of magnetic field lines using the combined field of an idealized flux-tube magnetic field and a one-dimensional fluctuating magnetic field. \citet{Tooprakai2007} studied the trapping of energetic particles using the same magnetic field model. However, the trapping of charged particles in the magnetic field generated by the two-component turbulence model has not been well explored by numerical simulations.

Although previous numerical simulations have successfully reproduced SEP dropouts \citep{Giacalone2000}, they assume \textit{ad hoc} pitch-angle scattering that is not realistic. Physically, the pitch-angle scattering of charged particles is due to small-scale magnetic fluctuations, which is not present in the \citet{Giacalone2000} model. The purpose of this study is to include the effect of small-scale magnetic turbulence and examine the propagation of SEPs in a turbulent magnetic field that has a power spectrum similar to that derived from observations. In this study, we use two different types of three-dimensional magnetic field turbulence models often used in studying the transport of energetic particles in space, i.e. the foot-point random motion model \citep{Jokipii1969,Giacalone2006} and two-component model \citep{Matthaeus1990}. The generated fluctuating magnetic field has a Kolmogorov-like power spectrum with wavelengths from just larger than the correlation scale, leading to large-scale field-line random walk, down through small scales that lead to resonant pitch-angle scattering of the particles. In Section $2$ we describe the magnetic turbulence models and numerical methods we use to study the propagation of energetic particles. The parameters used in the simulations are also listed. In Section $3$ we use test-particle simulations in the magnetic field generated from the two magnetic turbulence models to study the propagation of SEPs.  We show that dropouts during impulsive SEP events can be reproduced using the foot-point random motion model when the source region is small compared with the correlation scale. However, for the two-component model, we find that dropouts are not seen in the simulations for the parameters we use. Our order-of-magnitude estimate shows that in order for the intensity dropouts to form, particle scattering has to be infrequent, leading to a fairly large mean-free path. We discuss and summarize the results in Section $4$.

\section{Numerical Model \& Simulation}
In this study we consider the propagation of energetic particles from a spatially compact and instantaneous source in turbulent magnetic fields. We use two magnetic field turbulence models that capture the main observations of magnetic field fluctuation in the solar wind: the foot-point random motion model \citep[e.g.,][]{Jokipii1969,Jokipii1989,Giacalone2006} and two-component model \citep{Matthaeus1990,Bieber1996}. This section gives a mathematical description of the turbulent magnetic field models and the numerical method for integrating the trajectories of energetic charged particles.

\subsection{Turbulent Magnetic Fluctuations} \label{magnetic-model}
In a three-dimensional Cartesian geometry ($x, y, z$), a turbulent magnetic field can be expressed as
\begin{eqnarray} \textbf{B} &=& \textbf{B}_0 + \delta\textbf{B}\\
                            = &B_0& \hat{z}+ \delta B_x(x,y,z,t)\hat{x}+\delta B_y(x,y,z,t)\hat{y} + \delta B_z(x,y,z,t)\hat{z}. \nonumber \label{equation-mag}
\end{eqnarray}

\noindent This expression assumes a globally uniform background magnetic field $\textbf{B}_0$ in the $z$ direction and a fluctuating magnetic field component $\delta \textbf{B}$.

The two-component model is a quasi-static model for the wave-vector spectrum of magnetic fluctuation based on observations of the solar wind turbulence \citep{Matthaeus1990,Bieber1996}. In this model, the fluctuating magnetic field is expressed as the sum of two parts: a slab component $\delta \textbf{B}^{s} = (B_x^{s}(z), B_y^{s}(z), 0)$ and a two-dimensional component $\delta \textbf{B}^{2D} = (B_x^{2D}(x, y), B_y^{2D}(x, y), 0)$. The slab component is a one-dimensional fluctuating magnetic field with all wave vectors along the direction of the background magnetic field $\hat{z}$, and the two-dimensional component only consists of magnetic fluctuations with wave vectors along the transverse directions $\hat{x}$ and $\hat{y}$. It has been observed that magnetic field fluctuations have components with wave vectors nearly parallel or perpendicular to the magnetic field, with more wave power concentrated in the perpendicular directions (usually about $80\%$ in the solar wind) \citep{Bieber1996}. This model captures the anisotropic characteristic of the solar wind turbulence but neglects the turbulence component that propagates obliquely to the magnetic field $\textbf{B}_0$. 

Another often-used model for magnetic turbulence is based on the idea that magnetic fluctuations can be generated by foot-point random motions \citep{Jokipii1969,Jokipii1989,Giacalone2006}. One can consider a Cartesian geometry with the uniform magnetic field $\textbf{B}_0$ along the $z$ direction and the source surface lying in the $x$-$y$ plane at $z=0$. Since magnetic field lines are frozen in the surface velocity field, magnetic field fluctuations of the form in Equation \ref{equation-mag} can be produced by foot-point motions. We assume that the surface foot-point motion is described by $\textbf{v}_{fp}(x, y, t) = \nabla \times \Psi (x, y, t)$, where $\Psi$ is an arbitrary stream function. The fluctuating component of the magnetic field anywhere is given by
\begin{eqnarray}
\delta\textbf{B}^{fp} &=& \frac{B_0}{U}\textbf{v}_{fp}(x, y, t-z/U).
\end{eqnarray}

The magnetic field is assumed to have no dynamical variation but to be continuously dragged outward by a background fluid (the solar wind) with a convection speed $U$. When the magnetic field is evaluated at a certain time, it is fully three-dimensional with dependence on $x, y,$ and $z$.

In both of these two magnetic fluctuation models magnetic fields are variable in three spatial dimensions. As demonstrated by \citet{Jokipii1993} and \citet{Jones1998}, it is important to consider particle transport in a fully three-dimensional magnetic field because, in one- and two-dimensional fields, the particles adhere to the magnetic field lines on which they started their gyromotion due to the presence of at least one ignorable spatial coordinate. The magnetic fluctuations can be constructed using the random phase approximation \citep[e.g.,][]{Giacalone1999} and assuming a power spectrum of magnetic field fluctuations. This power spectrum can be determined from spacecraft observations \citep{Coleman1968,Bieber1993}. The slab component $\delta\textbf{B}^{s}$, two-dimensional component $\delta\textbf{B}^{2D}$, and fluctuating magnetic field produced by the foot-point random motion $\delta\textbf{B}^{fp}$ can be expressed as \citep{Giacalone1999,Giacalone2006}:
\begin{eqnarray}
&\delta &\textbf{B}^{s} = \sum^{N_{m}}_{n=1} A_n [\cos \alpha_n (\cos \phi_n \hat{x} +\sin \phi_n \hat{y}) \nonumber \\
& & + i\sin \alpha_n (-\sin\phi_n \hat{x} + \cos\phi_n \hat{y})] \nonumber\\
& &\times \exp(ik_nz + i\beta_n), \label{Bs}
\end{eqnarray}
\begin{eqnarray}
&\delta &\textbf{B}^{2D} = \sum^{N_{m}}_{n=1} A_n i(-\sin\phi_n \hat{x}+\cos\phi_n \hat{y}) \nonumber \\
&\times & \exp[i k_n (\cos\phi_n x+\sin \phi_n y) + i\beta_n], \label{B2d}
\end{eqnarray}
\begin{eqnarray}
&\delta &\textbf{B}^{fp} = (\hat{x}\frac{\partial}{\partial y} - \hat{y}\frac{\partial}{\partial x}) \\
&\times & \left[\sum^{N_{m}}_{n=1} \left(-\frac{1}{k_n}\right) A_n e^{ik_n(\cos \phi_n x + \sin \phi_n y)+i\omega_n(t-z/U)+i\beta_n}\right] \nonumber, \label{Bfp}
\end{eqnarray}

\noindent where $\beta_n$ is the phase of each wave mode, $A_n$ is its amplitude, $\omega_n$ is its frequency, $\alpha_n$ is the polarization angle, and $\phi_n$ determines the spatial direction of the $k$-vector in the $x$-$y$ plane. $\beta_n$, $\alpha_n$, and $\phi_n$ are random numbers between $0$ and $2\pi$. The frequency is taken to be $\omega_n = 0.1Uk_n$. This assumes that the Alfven speed is $0.1 U$, which is typical in the solar wind. All the forms of fluctuating magnetic fields satisfy the condition $\nabla \cdot \delta \textbf{B} = 0$.

The amplitude of the magnetic fluctuation at wave number $k_n$ is assumed to follow a Kolomogorov-like power law:
\begin{eqnarray}
A_n^2 = \sigma^2 \frac{\Delta V_n}{1 + (k_n L_c)^\gamma} \left[\sum^{N_m}_{n=1}\frac{\Delta V_n}{1+(k_n L_c)^\gamma}\right]^{-1},
\end{eqnarray}

\noindent where $\sigma^2$ is the magnetic wave variance and $\Delta V$ is a normalization factor. In one-dimensional, two-dimensional, and three-dimensional omnidirectional spectra,
$\Delta V_n = \Delta k_n$, $2\pi k_n \Delta k_n$, and $4\pi k_n^2 \Delta k_n$, and $\gamma = 5/3$, $8/3$, and $11/3$, respectively. A logarithmic spacing in wavenumbers $k_n$ is chosen so that $\Delta k_n/k_n$ is a constant.

It has been pointed out by \citet{Giacalone2006} that these two models are closely related and the two-component model can be reproduced using the foot-point random motion model by choosing a particular set of velocity field fluctuations. It should be noted that both of these two simplified models assume a quasi-static field that may not be appropriate for describing magnetic turbulence. We note that current sheets, which are known to exist in the solar wind
and are possibly related to turbulence, are not included in our kinematic model.
However, observations of dropouts show no correlation with current sheets  \citep{Mazur2000,Chollet2008b}.
Nevertheless, these two models are very useful in studying the transport of energetic charged particles in magnetic turbulence and explaining observations of SEP events. We also note that because our models use a Cartesian coordinate system with a constant solar wind speed and constant average magnetic field, the energetic particles do not undergo adiabatic cooling, nor adiabatic focusing.  These effects would arise in a
spherical  geometry, but are not included in our model.  Moreover, our models also assume that the magnetic-field turbulence variance does not change with distance.

In the simulations we generally use parameters similar to what is observed in the solar wind at $1$ AU. The minimum and maximum wavelengths $\lambda_{min}$ and $\lambda_{max}$ are taken to be $5 \times 10^{-5}$ AU and $1$ AU. The minimum wavelength is shorter than the resonant scale of particles with the lowest energy. We choose $\Delta k_n/k_n = 0.02$ and a total number of $N_m = 460$ wave modes are summed in the simulations.
The mean magnetic field $B_0$ is taken to be $5$ nT. The total variance of magnetic fluctuation $\sigma^2 = 0.3 B_0^2$. For the two-component model, the two-dimensional fluctuations contribute $80\%$ of the total magnetic power and the one-dimensional fluctuations contribute $20\%$ of the total magnetic power as suggested by \citet{Bieber1996}. The convection velocity of the solar wind $U$ is set to be $400$ km/s. The correlation length is assumed to be $L_c = 0.01$ AU. We note that these parameters, which are taken to not vary with distance from the Sun because of our assumed geometry, are based on observed values at 1 AU.  These parameters likely vary with distance from the Sun in reality.  However, we feel that some key parameters that govern the interplanetary scattering of particles may not vary significantly in the inner heliosphere.
 For example, Helios observations show that the correlation length only varies by a factor of about $5$ from $0.3$ AU to $1$ AU \citep[e.g.,][]{Bavassano1982a,Bruno2013}. The solar wind speed does not change significantly beyond the Alfven point \citep{McGregor2011}.  The ratio between the turbulence variance and the background magnetic field squared $\sigma^2/B_0^2$ only weakly depends on the radial distance \citep[e.g.,][]{Bavassano1982b}. Observations of SEP events show the scattering properties are similar between the inner heliosphere and at 1 AU, indicating the change of turbulence property does not cause strong variations in particle scattering \citep{Kallenrode1992,Kallenrode1993}.

\begin{figure}
\begin{tabular}{c}
\epsfig{file=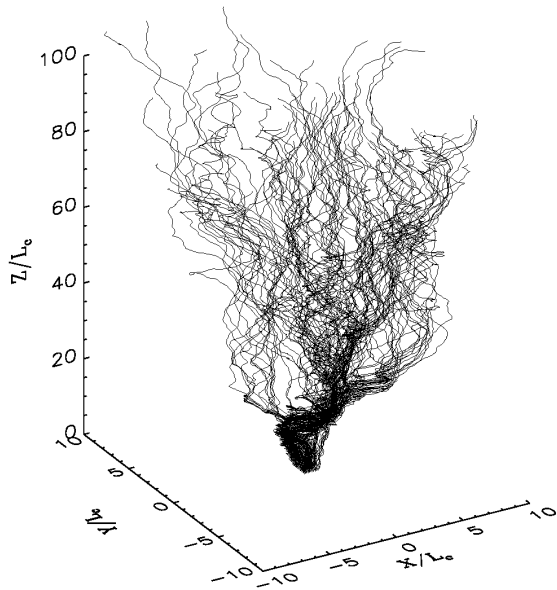,width=0.5\textwidth,clip=}
\end{tabular}
\caption[The turbulent magnetic field lines produced by the foot-point random motion model.]{The turbulent magnetic field lines produced by the foot-point random motion model originated from $-L_c<x<L_c$ and $-L_c<y<L_c$ at $t = 0$. See the text in Section \ref{magnetic-model} for description and parameters. \label{fig2-1}}
\end{figure}

In Figure \ref{fig2-1} we illustrate $100$ turbulent magnetic field lines originated from a surface region within $-L_c<x<L_c$ and $-L_c<y<L_c$ at $z = 0$ at time $t = 0$ produced by foot-point random motion. It is clear that the magnetic field lines meander on scales significantly greater than the correlation length. Field lines that start out from a compact source -- smaller than the correlation scale of turbulence -- meander considerably with increasing distance from the source in the $z$ direction. Figure $2$ shows the $x$ and $y$ components of the magnetic fields generated from two models at $z = 1$ AU. 

\begin{figure}
\begin{tabular}{c}
\epsfig{file=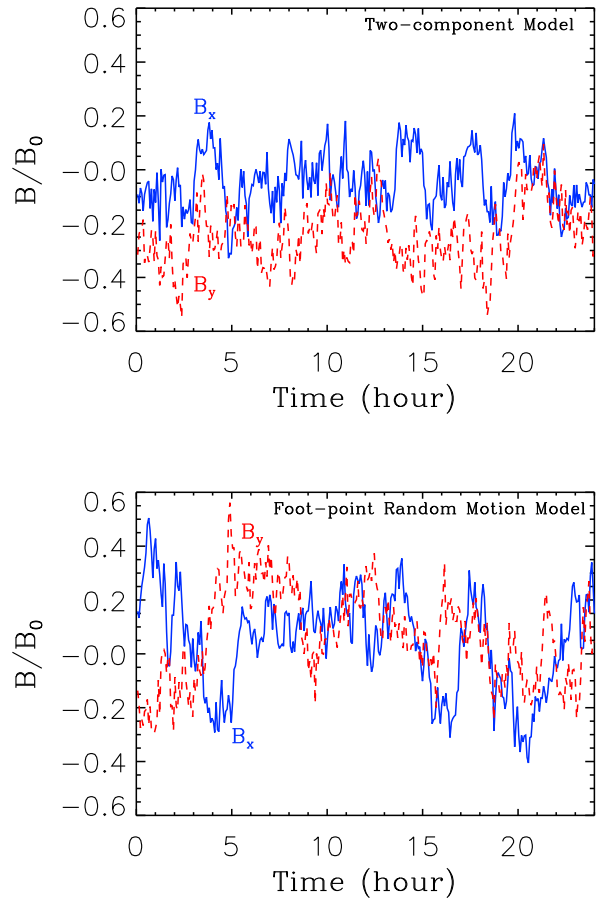,width=0.5\textwidth,clip=}
\end{tabular}
\caption[]{The $x$ and $y$ component of the turbulent magnetic field at $z = 1$ AU generated by the two-component model and foot-point random motion model. \label{magfield}}
\end{figure}

\subsection{Test-particle Simulations} \label{testparticle-model}

In order to study the propagation of energetic particles in the heliospheric magnetic field, we numerically integrate the trajectories of energetic particles in magnetic fields generated from the magnetic turbulence models described previously. At each time step, the magnetic field vector at the position of each particle is calculated from the magnetic turbulence models. The numerical technique used to integrate the trajectories of energetic particles is the Bulirsh-Stoer method, which is described in detail by \citet{Press1986}. It is highly accurate and conserves energy well. The algorithm uses an adjustable time-step method that is based on the evaluation of local truncation error. The time step is increased if the local truncation error is smaller than $10^{-6}$ for several consecutive time steps. In the case of no electric field, the energy of a single particle in the fluctuating magnetic field is conserved to a high degree with a less than $0.01\%$ change during the simulation. In our test-particle simulations the charged particles are released impulsively at $z = 0$ and their trajectories are numerically integrated until they reach boundaries at $z = 1.6$ AU and $z = -0.1$ AU. The spacecraft observations at $1$ AU are mimicked by collecting particles in windows of a size of $L_x \times L_y = 0.01$ AU $\times 0.01$ AU when the particles pass the windows at $z = 1$ AU. The size of the collection regions is about the same size as a correlation length. We have also used smaller collection windows (one fifth of a correlation length) to confirm the results presented in this paper. The collected particle energies and arrival times in each window are analysed as a spacecraft observation. The source regions are taken to be a circle at the $z = 0$ plane with a radius of $1$, much smaller than the correlation scale $R_{source} = 0.2 L_c$ and $2$, much larger than the correlation scale $R_{source} = 5 L_c$. A parameter list for different runs is given in Table 1. The energy for released particles ranges from $20$ keV to $10$ MeV. The velocity distribution of released particles is prescribed to follow a power law $f = f_0 v^{-4}$ with random pitch angles between $0^\circ$ and $90^\circ$. For each run, we release about $60$ million particles to get a sufficient number of particles reached the collection regions.

\section{Numerical Simulation of Small-scale Gradients in SEP Intensity Flux \label{chapter2-dropout}}

\begin{figure}
\begin{center}
\includegraphics[width=0.5\textwidth]{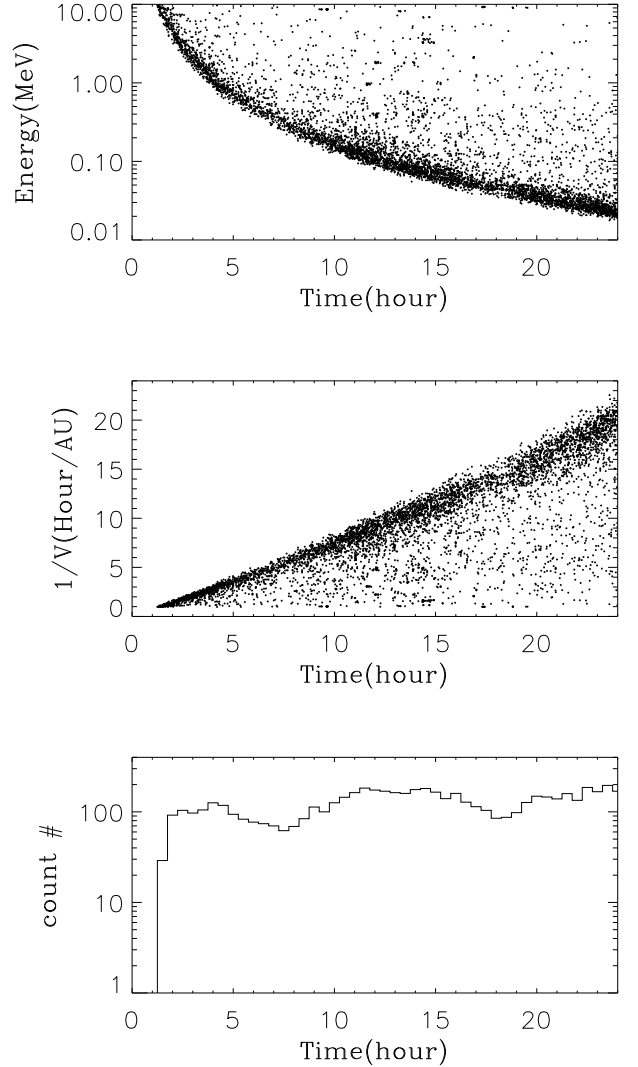}
\caption[An example of SEP event simulated using the foot-point random motion model for the case of a large source region.]{An example of SEP event simulated using the foot-point random motion model for the case of the large source region. \textit{Upper panel}: energy-time plot. \textit{Middle panel}: the inverse velocity $1/v$ versus the time after the release. \textit{Lower panel}: the count rate as a function of time in $30$-minute bins. The simulated event does not show any SEP dropouts. \label{fig-dropout-large}}
 \end{center}
 \end{figure}

\begin{figure}
\begin{center}
\includegraphics[width=0.5\textwidth]{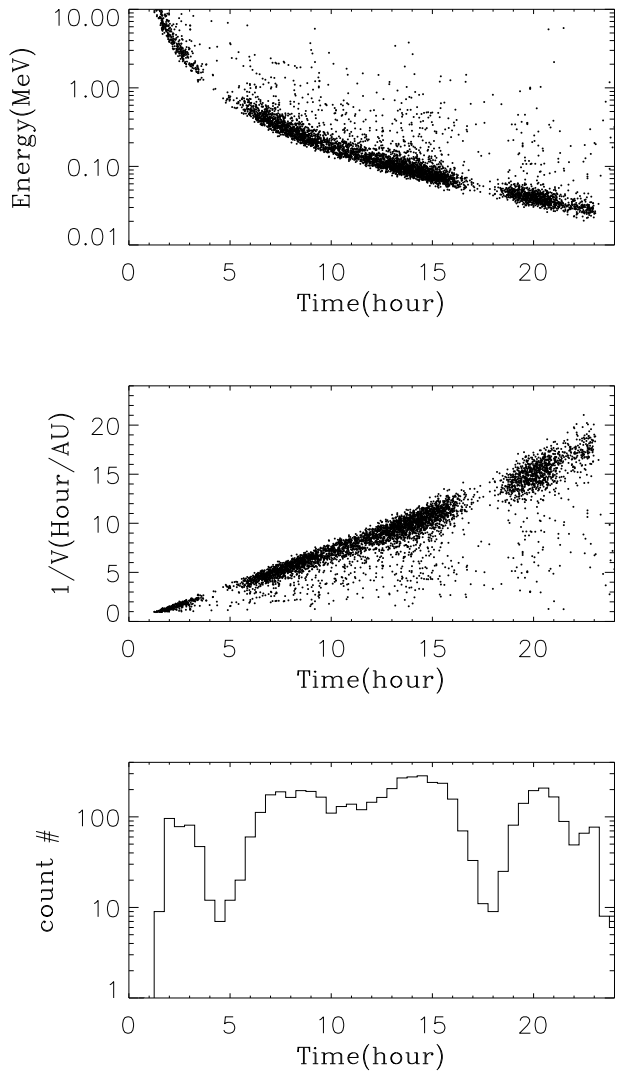}
\caption[An example of SEP dropouts simulated using the foot-point random motion model for the case of a small source region.]{An example of SEP dropouts simulated using the foot-point random motion model for the case of the small source region. \textit{Upper panel}: energy-time plot. \textit{Middle panel}: the inverse velocity $1/v$ versus the time after the release. \textit{Lower panel}: the count rate as a function of time in $30$-minute bins. This example clearly shows dropouts. \label{fig-dropout-small}}
 \end{center}
 \end{figure}

We use turbulent magnetic fields generated from the foot-point random motion model and two-component model described in Section 2 to study the SEP intensity dropouts observed by spacecraft such as ACE and \textit{Wind} \citep{Mazur2000,Chollet2008b}. Table 1 summarizes the simulation runs. In Figure \ref{fig-dropout-large} we show a simulated SEP event using the foot-point random motion model for the case of a large source region (Run $1$). The upper panel shows the energy-time plot and the middle panel shows the inverse velocity $1/v$ versus the time after the initial release. In the lower panel we show the count rate as a function of time in $30$-minute bins. One can see that in this case the simulated SEP event does not show any intensity dropout. The variation of flux is within a factor of two on the time scale of several hours after the initial onset. In the small source region case, dropouts can be frequently seen. An example is given in Figure \ref{fig-dropout-small}, which shows plots similar to Figure \ref{fig-dropout-large}, but for the small source region case (Run $2$). It is shown that two SEP dropouts can be clearly seen during $t = 3.7$ - $5.4$ hour and $t = 16.4$ - $18.3$ hour. The scatter plots show two clear gaps and the flux shows a more than one order-of-magnitude drop in about an hour. For the parameters we use, the convected distances for the dropouts are $2.4 \times 10^6$ km ($1.6 L_c$) and $2.7 \times 10^6$ km ($1.8 L_c$), respectively. These results are consistent with spacecraft observations \citep{Mazur2000,Chollet2008b}. The time intervals of these dropouts are typically several hours, which are similar to that observed in space \citep{Mazur2000}. These results are similar to those of \citet{Giacalone2000}, but unlike that study we have made no assumptions of \textit{ad hoc} scattereing. Instead, pitch-angle scattering in the present model is due to small-scale magnetic fluctuations.

\begin{figure}
\begin{center}
\includegraphics[width=0.5\textwidth]{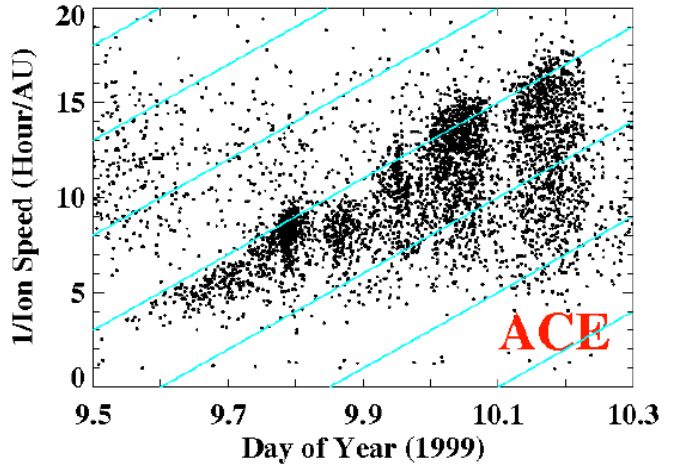}
\caption[An example of the observed SEP dropouts that show different path lengths observed by ACE/ULEIS detector.]{An example of the observed SEP dropouts that show different path lengths observed by ACE/ULEIS detector. Figure provided courtesy Dr. Joseph Mazur, Aerospace Corporation. }\label{flucedge}.
 \end{center}
 \end{figure}

\begin{figure}
\centering
\begin{tabular}{cc}
\epsfig{file=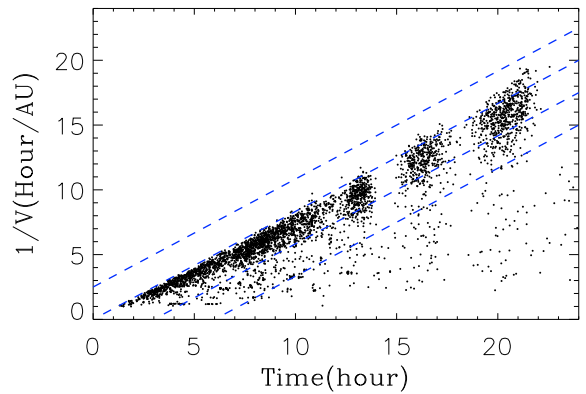,width=0.5\textwidth,clip=} \\
\end{tabular}
\caption{Examples of SEP dropouts produced from numerical simulations. The simulated SEP event shows different apparent path lengths. \label{figure-pathes}}
\end{figure}

SEPs accelerated close to the Sun can exhibit distinct velocity dispersions as they arrive at $1$ AU along different paths. Figure \ref{flucedge} shows an example of impulsive SEP events plotted as $1/v$ versus time (Figure provided courtesy Dr. Joseph Mazur). The observation was made by ACE/ULEIS detector in $1999$. It displays at least two distinct arrival times at $1$ AU, which indicates that particles follow at least two different field-line lengths. In our simulations, we also find that the apparent path lengths can have different values. An example of the simulated events is presented in Figure \ref{figure-pathes}. In this plot we use blue dashed lines as a reference, which represent particles travel along a field line with a length of $1.1$ AU and cosine pitch angle $\mu = v_\parallel/v = 1$. The time differences between the dashed lines are 3 hours.
 It can be seen that some particles collected at about $t = 16$ hour and $t = 20$ hour are above the second blue line, indicating that they arrived earlier than other particles. In addition, the slopes of velocity dispersions between dropouts are different, indicating particles travelling along field lines with distinct lengths.

\begin{figure}
\begin{center}
\includegraphics[width=0.5\textwidth]{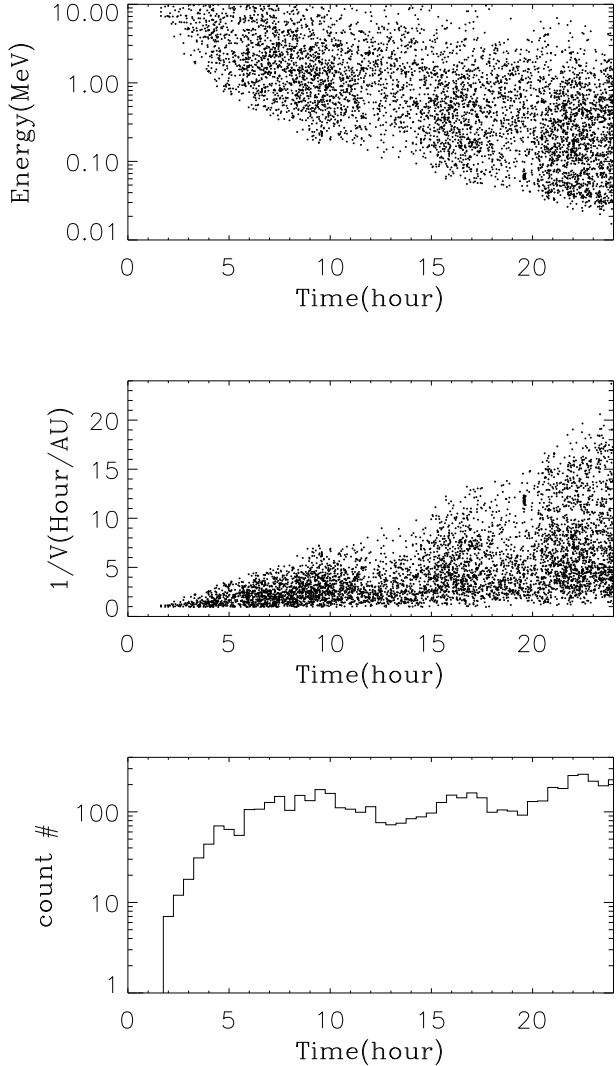}
\caption{An example of SEP event simulated using the two-component model for the case of a small source region. \textit{Upper panel}: energy-time plot. \textit{Middle panel}: the inverse velocity $1/v$ versus the time after the release. \textit{Lower panel}: the count rate as a function of time in $30$-minute bins. For the two-component model, the simulated events does not show any clear SEP dropouts for both small source region and large source region. \label{fig-dropout-2D}}
 \end{center}
 \end{figure}

\begin{figure}
\begin{center}
\includegraphics[width=0.5\textwidth]{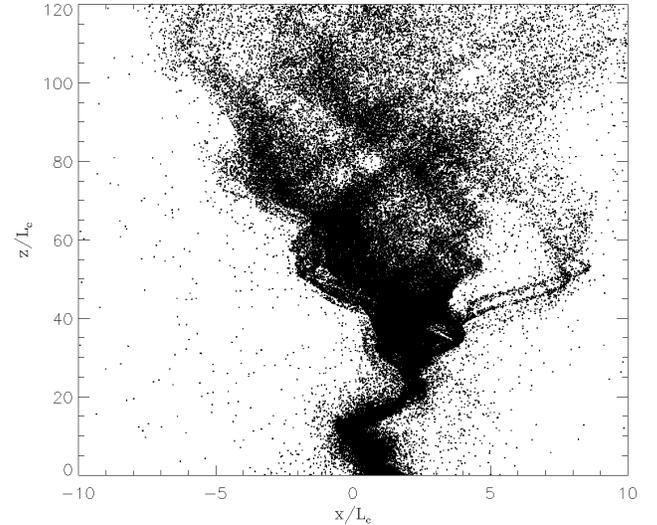}
\caption[The positions of energetic charged particles projected in $x - z$ plane at $t = 12$ hour. The results are from simulations using the foot-point random motion model.]{The positions of energetic charged particles projected in $x - z$ plane at $t = 14.4$ hour. The results are from the numerical simulations using the foot-point random motion model. It clearly shows that the particles follow the braiding magnetic field lines. }\label{fig-total-footpoint}
 \end{center}
 \end{figure}

\begin{figure}
\begin{center}
\includegraphics[width=0.5\textwidth]{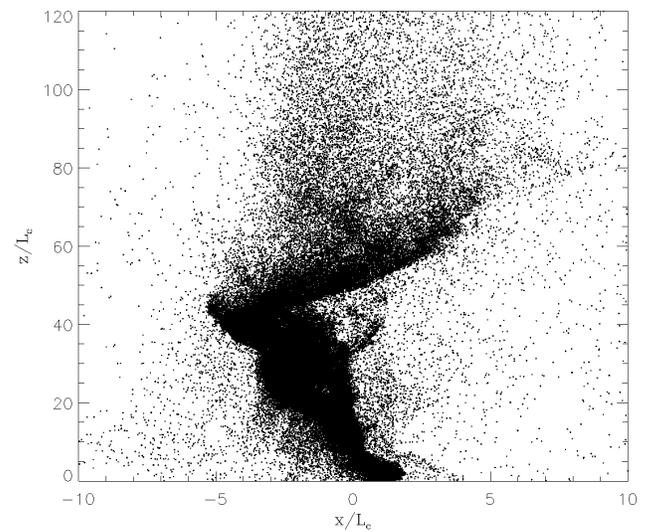}
\caption[The positions of energetic charged particles projected in $x - z$ plane at $t = 12$ hour. The results are from simulations using the two-component model.]{The positions of energetic charged particles projected in $x - z$ plane at $t = 14.4$ hour. The results are from the numerical simulations using the two-component model.}\label{fig-total-twocomponent}
 \end{center}
 \end{figure}

We have also attempted to use the two-component model to study SEP intensity dropouts. However, we did \textit{not} find any clear dropout in our simulations for either small or large source region. Figure \ref{fig-dropout-2D} shows a simulated SEP event using the two-component model when the source region is small (Run $4$). The particles in the energy-time plot and $1/v$-time plot show a more broadened distribution, indicating enhanced scattering compared with the cases for the foot-point random motion model. The simulated event has only some variations in flux within a factor of two during several hours. We have increased the number of particles and changed the size of the collection windows in the simulation and confirmed that those factors do not change this result.  To further resolve this issue, we have prepared two scatter plots that show the positions of about $400000$ energetic particles projected on the $x$ - $z$ plane at $t = 14.4$ hours after the initial release. The results are shown in Figure \ref{fig-total-footpoint} for the foot-point random motion model (Run 2) and in Figure \ref{fig-total-twocomponent} for the two-component model (Run 4). It can be clearly seen in Figure \ref{fig-total-footpoint} that the particles follow the braiding magnetic field lines, and therefore they are separated as the field lines meander in space. However, this feature is not clearly seen in Figure \ref{fig-total-twocomponent} for the two-component model. 

\begin{figure}
\begin{center}
\includegraphics[width=0.5\textwidth]{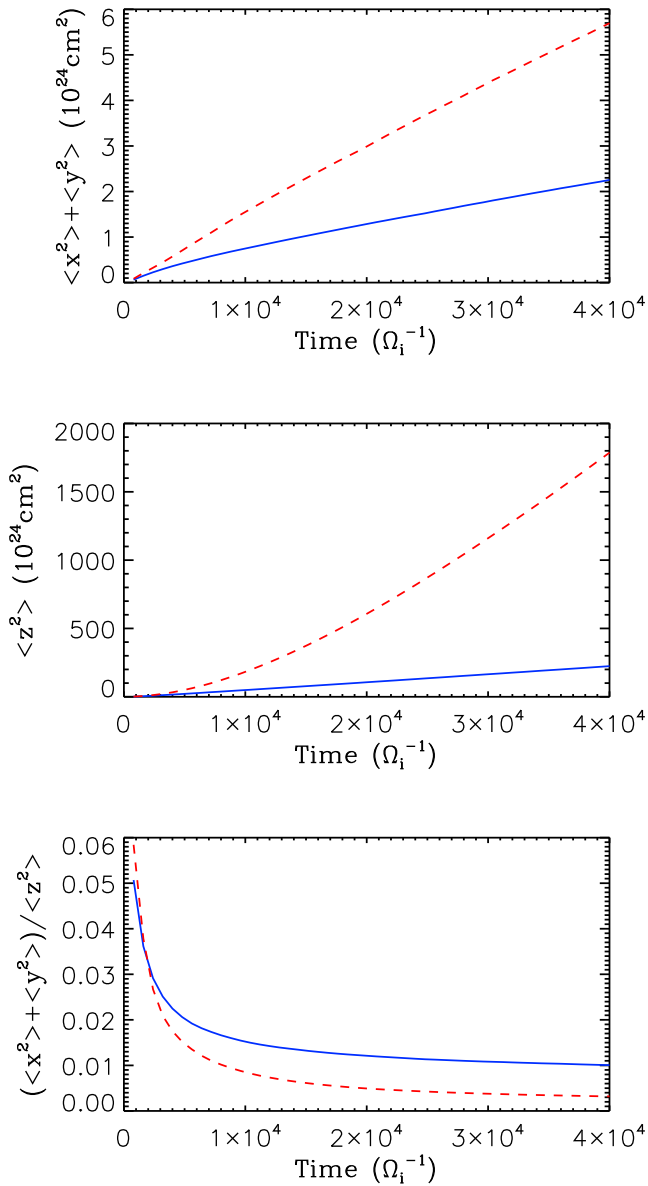}
\caption[Simulated diffusion coefficients of 1 MeV particles parallel and perpendicular to the mean magnetic field.]{Simulated diffusion coefficient of 1-MeV particles parallel (along the z-direction) and perpendicular (along the x- and y-direction) to the mean magnetic field. Blue lines represent results from the two-component model and red dashed lines represent results from the foot-point random motion model.}\label{fig-diffusion}
 \end{center}
 \end{figure}

A possible reason that dropouts are not seen in simulations using the two-component model is that the model contains a slab component that can more efficiently scatter energetic particles in pitch-angle. To demonstrate this, we measure diffusion coefficients of particle transport in the two models by implementing the technique used by \citet{Giacalone1999}. We use the definition of diffusion coefficients $\kappa_{\zeta \zeta} = $ $\langle \zeta^2 \rangle$/$2t$, where $\zeta$ is the spatial displacement at a given time $t$. We calculate the perpendicular and parallel diffusion coefficients for 1-MeV protons in the two turbulence models using the same parameters that were used in the simulation. The results are shown in Figure \ref{fig-diffusion}. Blue lines represent the displacements for particles propagating in the two-component model and red dashed lines represent results for foot-point random motion model. For the two-component model, the parallel and perpendicular coefficients are $\kappa_\parallel = 1.3 \times 10^{21}$ cm$^2$/s and $\kappa_\perp = 1.3 \times 10^{19}$ cm$^2$/s, respectively. For the foot-point random motion model, the parallel and perpendicular coefficients are $\kappa_\parallel = 1.07 \times 10^{22}$ cm$^2$/s and $\kappa_\perp = 3.4 \times 10^{19}$ cm$^2$/s, respectively. It is shown that the parallel diffusion coefficient for the two-component model is about one order of magnitude smaller than that for the foot-point random motion model, meaning that particles experience more scattering in the magnetic field generated from the two-component model. Because the inefficient pitch-angle scattering, particles follow field lines longer and the transfer from one field line to the next occurs less frequently. The calculation also shows a smaller ratio of $\kappa_\perp/\kappa_\parallel$ for the foot-point random motion model ($\kappa_\perp/\kappa_\parallel = 0.0032$) compared with that for the two-component model ($\kappa_\perp/\kappa_\parallel = 0.01$). Although drift motions due to the gradient and curvature of magnetic field can also cause particles travel off a field line, a small value of $\kappa_\perp/\kappa_\parallel$ indicates that the effect is fairly small. If large-scale field-line meandering happens to be the explanation for SEP dropouts, pitch-angle scattering due to small-scale scattering should be small so energetic particles can be mostly confined to their respective field lines. When the pitch-angle scattering is large, particles efficiently scatter off their original field lines and observers cannot see the intermittent intensity dropouts. 

We can quantitatively relate the mean-free path to the existence of intensity dropouts using an approach similar to that of \citet{Giacalone2000}.  
Consider charged particles, initially confined within some flux tube, that gradually leak off of the tube as they undergo pitch-angle scattering, causing them to be displaced normal to the local magnetic field by about one gyroradius each time they scatter.
 As the field line from which they
move initially along advects outward with the solar wind, an observer far from the source will see the particles decay from this field line due to their leakage off of it. The length scale of this decay is given by:
%
\begin{eqnarray}
L_{decay} = \sqrt{2\kappa_{\perp local}T}
\end{eqnarray}

\noindent where $\kappa_{\perp local}$ is the local cross-field diffusion coefficient, and $T = r/V_w$ is the time takes the fluid element -- to which the field is frozen -- that was initially associated with the injection of particles at the Sun to reach the observer. Here, $r$ is the distance between the source and the observer, and $V_w$ is the speed of the solar wind. In order for intensity dropouts to form, $L_{decay}$ must be less than the correlation scale $L_c$ of magnetic turbulence in the solar wind. We assume that locally the cross-field diffusion coefficient is given by:
%
\begin{eqnarray}
\kappa_{\perp local} = \kappa_\parallel (r_g/\lambda_\parallel)^2,
\end{eqnarray}

\noindent where $r_g = v/\Omega$ is the gyroradius of the energetic particle and $\Omega$ is its gyrofrequency, and $\lambda_\parallel$ is the parallel mean-free path. This is the so-called hard-sphere scattering approximation for cross-field diffusion and applies locally. Since $\kappa_\parallel = v\lambda_\parallel/3$, we find that if
\begin{eqnarray}
\lambda_\parallel \gg \frac{2v^3 r }{3L_c^2V_w\Omega^2},
\end{eqnarray}

\noindent then dropouts can occur. Using the parameters of our simulation, we find that for $1$ MeV protons, dropouts can result for $\lambda_{\parallel,1MeV} \gg 0.01$ AU. We also note that \citet{Chollet2011} directly measured $L_{decay}$ for some SEP events seen by ACE.  They found $L_{decay} = 0.001$ AU for $0.3$-$5$ MeV/nuc. ions.  Using the observed value of $L_{decay}$ in (7) and combining with (8), we solve for $\lambda_\parallel$ and find $\lambda_\parallel =  1$ AU.  This is consistent with our simulation results using the foot-point random motion model, but is larger than our results using the two-component model.  Thus, we suggest that the reason there are no dropouts in the two-component model is that the scattering mean-free path is too short and the dropouts, which may be present closer to the Sun, are filled in by $1$ AU.

To further test the effect of pitch-angle scattering, we make one more numerical simulation using the two-component model. In this case the smallest wavelength is taken to be $2 \times 10^{-3}$ AU, which is larger than the resonant scale of the protons of highest energy ($10$ MeV) in our simulation. Thus, we essentially remove small-scale resonant scattering, which have the effect of increasing the mean-free path. We find that in this case particles indeed follow and are separated by the braiding magnetic field lines and dropouts can form. This indicates that if pitch-angle scattering is infrequent, small-scale sharp gradients in SEP intensity flux can form. In Figure \ref{fig-dropout-chop} we show a simulated event using the two-component model with a minimum wavelength of $2 \times 10^{-3}$ AU. The event shows dropouts similar to the case using the foot-point random motion model. Two main drops in flux can be readily seen.

\begin{figure}
\begin{center}
\includegraphics[width=0.5\textwidth]{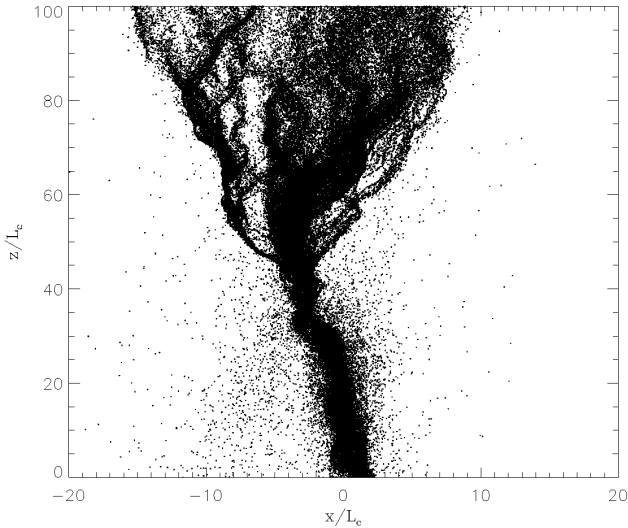}
\caption{The positions of energetic charged particles projected in $x - z$ plane at $t = 14.4$ hour. The results are from the numerical simulations using the two-component model but the smallest wavelength is taken to be $2 \times 10^{-3}$ AU.}\label{fig-large}
 \end{center}
 \end{figure}
 
 \begin{figure}
\begin{center}
\includegraphics[width=0.5\textwidth]{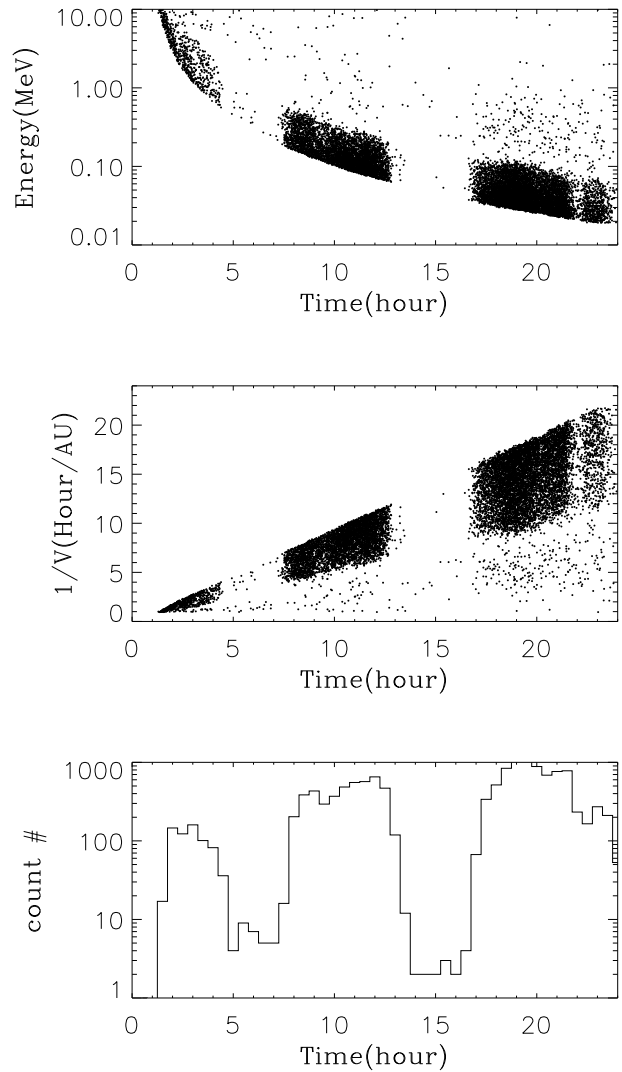}
\caption{An example of SEP dropouts simulated using the two-component model with a minimum wavelength $\lambda_{min} = 2\times 10^{-3}$ AU. \textit{Upper panel}: energy-time plot. \textit{Middle panel}: the inverse velocity $1/v$ versus the time after the release. \textit{Lower panel}: the count rate as a function of time in $30$-minute bins. This example clearly shows dropouts. \label{fig-dropout-chop}}
 \end{center}
 \end{figure}

\section{Discussion and Summary \label{chapter2-summary}}

In this paper, we studied small-scale gradients in the intensity of SEPs when energetic particles propagate in the heliospheric magnetic field. We numerically integrated the trajectories of energetic charged particles in the turbulent magnetic field generated from the commonly used magnetic turbulence models, i.e. the foot-point random motion model \citep{Jokipii1969,Giacalone2006} and two-component model \citep{Matthaeus1990}. The turbulence models include a Kolmogorov-like magnetic field power spectrum containing a broad range of spatial scales from those that lead to large-scale field-line random walk to small scales leading to resonant pitch-angle scattering of energetic particles. The observations of SEP events were simulated by collecting charged particles that reach $1$ AU much as a spacecraft detector would.

We have reproduced SEP dropouts in the numerical simulation using the foot-point random motion model, assuming that the radius of the SEP source region is smaller than the correlation scale of magnetic turbulence. The intervals of these dropouts are typically several hours, similar to the time scale of dropouts observed in space. The velocity dispersions of simulated SEP events appear to have distinct path lengths, which indicates that the energetic particles travel along field lines with different lengths. We have also attempted to use the two-component model to simulate dropouts in SEP intensity. However, we did not find any clear evidence of intensity dropouts in the simulation. This is probably because particle scattering is more efficient in the two-component model compared with that in the foot-point random motion model. We have demonstrated this by calculating diffusion coefficients in the simulations. This explanation is also supported by recent observational analysis by \citet{Chollet2011}. They inferred the intensity-fall-off lengths at the edges of the dropouts using ACE/ULEIS data and showed that energetic particles rarely scatter off a magnetic field line during the propagation in interplanetary turbulence. We showed that the parallel mean-free path inferred from this observation is consistent with that calculated using the foot-point random motion model, but about one order of magnitude larger than that in the two-component model.

Determining the value of large-scale diffusion coefficients of charged particles propagating in turbulent magnetic fields has been a long-standing issue for many decades. It is well-known that the mean-free paths inferred from SEP observations are usually much longer than those derived from quasi-linear theory \citep{Palmer1982,Bieber1994,He2012}. It has been proposed that the anisotropy of magnetic turbulence can strongly influence the diffusion coefficient parallel to the magnetic field \citep{Bieber1996}. However,  it was later found that the diffusion coefficient only has a weak dependence on the anisotropy \citep{Giacalone1999,Qin2006}. Here we showed that small-scale gradients in the intensity of energetic particles during impulsive SEP events can provide a strong constraint on the mean-free path of charged particles. As energetic particles propagate from the source close to the Sun to $1$ AU, pitch-angle scattering caused by small-scale magnetic fluctuations has to be infrequent so that energetic particles can be largely confined to their original field lines. Otherwise charged particles can effectively move off field lines and dropouts cannot form. 

It is worth noting that the magnetic turbulence models we used have a fixed set of parameters like magnitude of magnetic field and turbulence variance. This is needed since in our model magnetic field fluctuations are generated in a Cartesian geometry. The parameters can actually vary from close to the Sun to $1$ AU. Although previous studies have shown this variation does not significantly change the scattering of particles \citep{Kallenrode1992,Kallenrode1993}. This effect may need to be considered in the future.

It is also worth noting that we did not observe a significant effect of particle trapping on magnetic field lines with a restricted transverse displacement as proposed by \citet{Ruffolo2003} and \citet{Chuychai2007}. \citet{Tooprakai2007} have studied numerically the trapping process for charged particles in magnetic fluctuations. The difference between our simulations and their simulations is that we directly used the two-component model as proposed by \citet{Matthaeus1990} and \citet{Bieber1996}, whereas \citet{Tooprakai2007} used a combined field of a idealized two-dimensional flux-tube magnetic field and a one-dimensional fluctuating magnetic field.

\section*{Acknowledgement}
F.G. benefited from the conversations with Randy Jokipii, Jozsef Kota,  Federico Fraschetti, Andrey Beresnyak, Gang Qin, Joseph Mazur, and William Matthaeus. This work was supported by NASA under grant NNX11AO64G and by NSF under grant AGS1154223. Part of the computational resource supporting this work were provided by the NASA High-End Computing (HEC) Program through the NASA Advanced Supercomputing (NAS) Division at Ames Research Center.

\singlespace
\bibliographystyle{apj}

\begin{thebibliography}{35}
\expandafter\ifx\csname natexlab\endcsname\relax\def\natexlab#1{#1}\fi

\bibitem[Bavassano et al.(1982a)]{Bavassano1982a} Bavassano, B., 
Dobrowolny, M., Mariani, F., \& Ness, N.~F.\ 1982a, \jgr, 87, 3617

\bibitem[Bavassano et al.(1982b)]{Bavassano1982b} Bavassano, B., 
Dobrowolny, M., Fanfoni, G., Mariani, F., 
\& Ness, N.~F.\ 1982b, \solphys, 78, 373 

\bibitem[Bieber et al.(1993)]{Bieber1993} Bieber, J.~W., Chen, J., 
Matthaeus, W.~H., Smith, C.~W., \& Pomerantz, M.~A.\ 1993, \jgr, 98, 3585 

\bibitem[{{Bieber} {et~al.}(1994){Bieber}, {Matthaeus}, {Smith}, {Wanner},
  {Kallenrode}, \& {Wibberenz}}]{Bieber1994}
{Bieber}, J.~W., {Matthaeus}, W.~H., {Smith}, C.~W., {Wanner}, W.,
  {Kallenrode}, M.-B., \& {Wibberenz}, G. 1994, \apj, 420, 294

\bibitem[{{Bieber} {et~al.}(1996){Bieber}, {Wanner}, \&
  {Matthaeus}}]{Bieber1996}
{Bieber}, J.~W., {Wanner}, W., \& {Matthaeus}, W.~H. 1996, \jgr, 101, 2511

\bibitem[Bruno 
\& Carbone(2013)]{Bruno2013} Bruno, R., \& Carbone, V.\ 2013, Living Reviews in Solar Physics, 10, 2 

\bibitem[{{Chandran}(2000)}]{Chandran2000}
{Chandran}, B.~D.~G. 2000, Physical Review Letters, 85, 4656

\bibitem[{{Chollet}(2008)}]{Chollet2008a}
{Chollet}, E.~E. 2008, PhD thesis, The University of Arizona

\bibitem[{{Chollet} \& {Giacalone}(2008)}]{Chollet2008b}
{Chollet}, E.~E., \& {Giacalone}, J. 2008, \apj, 688, 1368

\bibitem[{{Chollet} \& {Giacalone}(2011)}]{Chollet2011}
---. 2011, \apj, 728, 64

\bibitem[{{Chollet} {et~al.}(2007){Chollet}, {Giacalone}, {Mazur}, \& {Al
  Dayeh}}]{Chollet2007}
{Chollet}, E.~E., {Giacalone}, J., {Mazur}, J.~E., \& {Al Dayeh}, M. 2007,
  \apj, 669, 615

\bibitem[{{Chuychai} {et~al.}(2007){Chuychai}, {Ruffolo}, {Matthaeus}, \&
  {Meechai}}]{Chuychai2007}
{Chuychai}, P., {Ruffolo}, D., {Matthaeus}, W.~H., \& {Meechai}, J. 2007, \apj,
  659, 1761
  
\bibitem[Coleman(1968)]{Coleman1968} Coleman, P.~J., Jr.\ 1968, 
\apj, 153, 371

\bibitem[{{Dresing} {et~al.}(2012){Dresing}, {G{\'o}mez-Herrero}, {Klassen},
  {Heber}, {Kartavykh}, \& {Dr{\"o}ge}}]{Dresing2012}
{Dresing}, N., {G{\'o}mez-Herrero}, R., {Klassen}, A., {Heber}, B.,
  {Kartavykh}, Y., \& {Dr{\"o}ge}, W. 2012, Solar Physics, 281, 281

\bibitem[{{Dwyer} {et~al.}(1997){Dwyer}, {Mason}, {Mazur}, {Jokipii}, {von
  Rosenvinge}, \& {Lepping}}]{Dwyer1997}
{Dwyer}, J.~R., {Mason}, G.~M., {Mazur}, J.~E., {Jokipii}, J.~R., {von
  Rosenvinge}, T.~T., \& {Lepping}, R.~P. 1997, \apjl, 490, L115

\bibitem[{{Giacalone} \& {Jokipii}(1999)}]{Giacalone1999}
{Giacalone}, J., \& {Jokipii}, J.~R. 1999, \apj, 520, 204

\bibitem[{{Giacalone} \& {Jokipii}(2012)}]{Giacalone2012}
---. 2012, \apjl, 751, L33

\bibitem[{{Giacalone} {et~al.}(2006){Giacalone}, {Jokipii}, \&
  {Matthaeus}}]{Giacalone2006}
{Giacalone}, J., {Jokipii}, J.~R., \& {Matthaeus}, W.~H. 2006, \apjl, 641, L61

\bibitem[{{Giacalone} {et~al.}(2000){Giacalone}, {Jokipii}, \&
  {Mazur}}]{Giacalone2000}
{Giacalone}, J., {Jokipii}, J.~R., \& {Mazur}, J.~E. 2000, \apjl, 532, L75

\bibitem[{{He} \& {Wan}(2012)}]{He2012}
{He}, H.-Q., \& {Wan}, W. 2012, \apj, 747, 38

\bibitem[{{Jokipii}(1966)}]{Jokipii1966}
{Jokipii}, J.~R. 1966, \apj, 146, 480

\bibitem[{{Jokipii} \& {Kota}(1989)}]{Jokipii1989}
{Jokipii}, J.~R., \& {Kota}, J. 1989, \grl, 16, 1

\bibitem[{{Jokipii} {et~al.}(1993){Jokipii}, {Kota}, \&
  {Giacalone}}]{Jokipii1993}
{Jokipii}, J.~R., {Kota}, J., \& {Giacalone}, J. 1993, \grl, 20, 1759

\bibitem[{{Jokipii} \& {Parker}(1969)}]{Jokipii1969}
{Jokipii}, J.~R., \& {Parker}, E.~N. 1969, \apj, 155, 777

\bibitem[{{Jones} {et~al.}(1998){Jones}, {Jokipii}, \& {Baring}}]{Jones1998}
{Jones}, F.~C., {Jokipii}, J.~R., \& {Baring}, M.~G. 1998, \apj, 509, 238

\bibitem[Kallenrode et al.(1992)]{Kallenrode1992} Kallenrode, M.-B., 
Wibberenz, G., \& Hucke, S.\ 1992, \apj, 394, 351 

\bibitem[Kallenrode(1993)]{Kallenrode1993} Kallenrode, M.-B.\ 1993, 
\jgr, 98, 19037

\bibitem[{{Matthaeus} {et~al.}(1990){Matthaeus}, {Goldstein}, \&
  {Roberts}}]{Matthaeus1990}
{Matthaeus}, W.~H., {Goldstein}, M.~L., \& {Roberts}, D.~A. 1990, \jgr, 95,
  20673

\bibitem[{{Matthaeus} {et~al.}(2003){Matthaeus}, {Qin}, {Bieber}, \&
  {Zank}}]{Mattaeus2003}
{Matthaeus}, W.~H., {Qin}, G., {Bieber}, J.~W., \& {Zank}, G.~P. 2003, \apjl,
  590, L53

\bibitem[{{Mazur} {et~al.}(2000){Mazur}, {Mason}, {Dwyer}, {Giacalone},
  {Jokipii}, \& {Stone}}]{Mazur2000}
{Mazur}, J.~E., {Mason}, G.~M., {Dwyer}, J.~R., {Giacalone}, J., {Jokipii},
  J.~R., \& {Stone}, E.~C. 2000, \apjl, 532, L79
  
  \bibitem[McGregor et al.(2011)]{McGregor2011} McGregor, S.~L., 
Hughes, W.~J., Arge, C.~N., Odstrcil, D., 
\& Schwadron, N.~A.\ 2011, Journal of Geophysical Research (Space Physics), 116, 3106 


\bibitem[{{Palmer}(1982)}]{Palmer1982}
{Palmer}, I.~D. 1982, Reviews of Geophysics and Space Physics, 20, 335

\bibitem[{{Parker}(1965)}]{Parker1965}
{Parker}, E.~N. 1965, \planss, 13, 9

\bibitem[{{Press} {et~al.}(1986){Press}, {Flannery}, \&
  {Teukolsky}}]{Press1986}
{Press}, W.~H., {Flannery}, B.~P., \& {Teukolsky}, S.~A. 1986, {Numerical
  recipes. The art of scientific computing}, ed. {Press, W.~H., Flannery,
  B.~P., \& Teukolsky, S.~A.}

\bibitem[{{Qin} {et~al.}(2002){Qin}, {Matthaeus}, \& {Bieber}}]{Qin2002}
{Qin}, G., {Matthaeus}, W.~H., \& {Bieber}, J.~W. 2002, \apjl, 578, L117

\bibitem[{{Qin} {et~al.}(2006){Qin}, {Matthaeus}, \& {Bieber}}]{Qin2006}
---. 2006, \apjl, 640, L103



\bibitem[{{Roelof} {et~al.}(1983){Roelof}, {Decker}, \&
  {Krimigis}}]{Roelof1983}
{Roelof}, E.~C., {Decker}, R.~B., \& {Krimigis}, S.~M. 1983, \jgr, 88, 9889

\bibitem[{{Ruffolo} {et~al.}(2003){Ruffolo}, {Matthaeus}, \&
  {Chuychai}}]{Ruffolo2003}
{Ruffolo}, D., {Matthaeus}, W.~H., \& {Chuychai}, P. 2003, \apjl, 597, L169

\bibitem[{{Tooprakai} {et~al.}(2007){Tooprakai}, {Chuychai}, {Minnie},
  {Ruffolo}, {Bieber}, \& {Matthaeus}}]{Tooprakai2007}
{Tooprakai}, P., {Chuychai}, P., {Minnie}, J., {Ruffolo}, D., {Bieber}, J.~W.,
  \& {Matthaeus}, W.~H. 2007, \grl, 34, 17105
  
\bibitem[Wicks et al.(2010)]{Wicks2010} Wicks, R.~T., Owens, 
M.~J., \& Horbury, T.~S.\ 2010, \solphys, 262, 191 

\bibitem[{{Wiedenbeck} {et~al.}(2013){Wiedenbeck}, {Mason}, {Cohen}, {Nitta},
  {G{\'o}mez-Herrero}, \& {Haggerty}}]{Wiedenbeck2013}
{Wiedenbeck}, M.~E., {Mason}, G.~M., {Cohen}, C.~M.~S., {Nitta}, N.~V.,
  {G{\'o}mez-Herrero}, R., \& {Haggerty}, D.~K. 2013, \apj, 762, 54

\bibitem[{{Wiedenbeck} {et~al.}(2010){Wiedenbeck}, {Mason},
  {G{\'o}mez-Herrero}, {Haggerty}, {Nitta}, {Cohen}, {Chollet}, {Cummings},
  {Leske}, {Mewaldt}, {Stone}, {von Rosenvinge}, {M{\"u}ller-Mellin}, {Desai},
  \& {Mall}}]{Wiedenbeck2010}
{Wiedenbeck}, M.~E., {et~al.} 2010, Twelfth International Solar Wind
  Conference, 1216, 621

\bibitem[{{Yan} \& {Lazarian}(2002)}]{Yan2002}
{Yan}, H., \& {Lazarian}, A. 2002, Physical Review Letters, 89, B1102

\bibitem[{{Zhang} {et~al.}(2003){Zhang}, {Jokipii}, \& {McKibben}}]{Zhang2003}
{Zhang}, M., {Jokipii}, J.~R., \& {McKibben}, R.~B. 2003, \apj, 595, 493

\end{thebibliography}

 \begin{table}
\begin{tabular*}
{0.7\textwidth}{ccccc}
\hline 
Run &          model         & source radius ($L_c$)   &    minimum wavelength (AU)   & dropouts?   \\
\hline
1   & field line random walk &   $5.0$   &   $5 \times 10^{-5} $  & N \\
2   & field line random walk &   $0.2$    &   $5 \times 10^{-5} $  & Y \\
3   & two component          &   $5.0$    &   $5 \times 10^{-5} $  & N   \\
4   & two component          &   $0.2$    &   $5 \times 10^{-5} $  & N \\
5   & two component          &   $0.2$    &   $2 \times 10^{-3} $  & Y \\

 \hline
\end{tabular*}
 \caption{List of simulation runs. $L_c$ is taken to be $0.01$ AU.}
 \label{table1}
\end{table}

\clearpage

\end{document}